\documentclass[hypertexnames=false,twocolumn,aps,superscriptaddress,prl]{revtex4-2}
\usepackage{array, makecell}

\usepackage{fancyref}
\usepackage{hyperref}
\usepackage{dcolumn}
\usepackage{bm}
\usepackage{tikz}
\usepackage{pgfplots}
\pgfplotsset{compat=1.18}
\tikzset{>=latex}
\usetikzlibrary{quantikz}
\usepackage{amsfonts}

\usepackage{booktabs}

\usepackage[mode=buildnew]{standalone}

\usepackage{lipsum}
\usepackage{graphicx} 
\usepackage{standalone}
\usepackage[caption=false]{subfig}
\usepackage[percent]{overpic}
\usepackage{verbatim}
\usepackage{tcolorbox}

\usepackage{physics} 
\usepackage{enumerate}
\usepackage{dsfont}
\usepackage{xcolor}

\usepackage[normalem]{ulem}
\usepackage[capitalise]{cleveref}

\usepackage{lipsum}

\begin{document}

\title{Loss-biased fault-tolerant quantum error correction}

\author{Laura Pecorari}
\affiliation{University of Strasbourg and CNRS, CESQ and ISIS (UMR 7006), aQCess, 67000 Strasbourg, France}

\author{Gavin K. Brennen}
\affiliation{Center for Engineered Quantum Systems, School of Mathematical and Physical Sciences, Macquarie University, 2109 NSW, Australia}
\affiliation{BTQ Technologies, 16-104 555 Burrard Street, Vancouver, British Columbia, Canada V7X 1M8}

\author{Stanimir S. Kondov}
\affiliation{University of Strasbourg and CNRS, CESQ and ISIS (UMR 7006), aQCess, 67000 Strasbourg, France}

\author{Guido Pupillo}
\affiliation{University of Strasbourg and CNRS, CESQ and ISIS (UMR 7006), aQCess, 67000 Strasbourg, France}

\date{\today}

\begin{abstract}
We investigate the limits of quantum error correction (QEC) in neutral-atom processors approaching high-fidelity gates and fast cycle times. We show that shorter QEC cycles amplify platform-specific errors, notably Rydberg excitation hopping, and hinder decay of residual Rydberg population, leading to non-Markovian correlated errors that degrade logical performance. To address this, we introduce loss biasing, where spurious Rydberg excitations are rapidly converted into atom loss via mid-circuit ionization, transforming errors into erasure-like noise and suppressing their propagation. Loss biasing restores the fault-tolerant logical error scaling for intra-cycle Pauli errors; furthermore, we argue that when supported with loss-aware decoding, it can achieve the optimal scaling of erasures while enabling shorter QEC cycles with reduced hardware overhead. We outline an implementation using fast autoionization in alkaline-earth(-like) atoms, establishing loss biasing as a practical route toward fault-tolerant quantum computing with sub-millisecond QEC cycles.
\end{abstract}

\maketitle
Neutral atoms in optical tweezers have emerged as a leading platform for quantum information processing, combining flexible connectivity, intrinsic scalability, and high-fidelity operations \cite{Bluvstein_2022,Manetsch_2025,Evered_2023,Scholl_2023,Ma_2023,senoo2025highfidelityentanglementcoherentmultiqubit} -- key ingredients for efficient quantum error correction (QEC) \cite{gottesman1997stabilizercodesquantumerror}. 
Two-qubit CZ gates based on the Rydberg-blockade mechanism \cite{Jandura_2022} now reach infidelities at the $10^{-3}$ level on sub-$\mu$s timescales, with a significant fraction of errors arising from leakage outside the computational subspace ($\sim$40\%) \cite{Bluvstein_2025}. Given sufficient time (typically 100's of $\mu$s), residual Rydberg population generated by gate errors is likely to convert into qubit loss: either through decay into low-lying non-computational states or through atom loss. Processor architectures which rely on qubit rearrangement between entangling gates naturally accommodate for this process, as QEC cycles — comprising multiple CZ gates interleaved with atom shuttling — are limited by rearrangement speed ($\sim0.55$ m/s \cite{Bluvstein_2022}). Suitable error-correction and qubit replenishment protocols have recently been demonstrated in both theory and experiment \cite{Chow_2024,Perrin_2025,ycwc-3myc,Bluvstein_2025,liu2026achievingoptimaldistanceatomlosscorrection,perrin2026correlatedatomlossresource}. However, next-generation architectures aim to significantly tighten CZ timing and decrease cycle times in the pursuit of faster logical execution.

In this work, we quantitatively show that shrinking QEC cycle times introduce two key challenges for quantum error correction with neutral atoms. 
First, as gate fidelities improve and inter-gate delays decrease, platform-specific error mechanisms become dominant; in neutral-atom systems, this is primarily \emph{Rydberg hopping}, i.e., the transfer of a Rydberg excitation during two-qubit gates \cite{jandura2024surfacecodestabilizermeasurements}. Second, shorter cycles reduce the time available for residual Rydberg excitations to decay, leading to correlated error clusters that degrade logical performance and fault tolerance. These effects are particularly relevant for quasi-static architectures that minimize atom shuttling \cite{rines2025demonstrationlogicalarchitectureuniting}, as well as for next-generation high-speed platforms.
To address these challenges, we introduce \emph{loss biasing}: the rapid conversion of spurious Rydberg excitations into atom loss via mid-circuit ionization. The resulting noise is biased toward qubit loss, which, unlike conventional bias-preserving schemes \cite{Puri_2020}, does not require active stabilization, as a loss effectively acts as a reset mechanism (or as a persistent qubit in $\ket{0}$) that decouples the atom from subsequent Rydberg gates pulses. 
When combined with leakage reduction units and loss-aware decoding, such loss events can recover the optimal logical error scaling of $\sim d$ with code distance $d$, analogous to erasure errors \cite{Perrin_2025,ycwc-3myc,liu2026achievingoptimaldistanceatomlosscorrection,perrin2026correlatedatomlossresource}, even when atom replenishment is delayed at the end of the QEC cycle.
Loss biasing can thus achieve QEC performance comparable to state-of-the-art erasure-conversion protocols \cite{Wu_2022}, which require ionization, detection, and mid-circuit qubit reloading, but with reduced cycle time at the cost of moderately increased decoding complexity. Thus, this approach provides a fast, hardware-efficient mechanism to suppress hopping-induced errors and restore fault-tolerant scaling.  We outline a concrete implementation using alkaline-earth(-like) atoms, where fast, high-fidelity loss conversion can be realized via autoionization, providing a practical route toward sub-ms QEC cycles.

We consider a distance-$d$ rotated surface code \cite{PhysRevA.76.012305} [Fig.~\ref{fig:figure1}(a)] that uses $d^2$ data qubits and $d^2-1$ ancilla qubits to encode a single logical qubit. We model each atom as a three-level system, with computational basis states $\ket{0}$, $\ket{1}$ and Rydberg state $\ket{r}$ [Fig.~\ref{fig:scalings}(a)]. The $5$-atom plaquette Hamiltonian reads
\begin{align}
\label{eq:hamiltonian}
    \begin{split}
    H = \sum_{i=0}^4\Delta E_i\ket{1_i}\bra{1_i}&+\sum_{ij=0}^{4} B_{ij}\ket{r_ir_j}\bra{r_ir_j} \\
    &+\left(\sum_{i=0}^{4} \frac{\Omega_i(t)}{2}\ket{r_i}\bra{1_i} + \mathrm{h.c.}\right)
    \end{split}
\end{align}
where $\Delta E_i$ is the energy splitting between the computational qubit states of atom $i$, $B_{ij}$ is the interaction strength between atoms $i$ and $j$ excited to their Rydberg state $\ket{r_ir_j}$, $\Omega_i(t)$ is the complex time-dependent Rabi frequency of the driving laser coupling $\ket{1_i}$ and $\ket{r_i}$ states of atom $i$. In the following, we consider time-optimal symmetric pulses \cite{Jandura_2022} for each gate, $\Omega_i(t)\equiv\Omega(t)\,\forall i$, and assume only data-ancilla blockade with either $B_{ij}=0$ (perfect absence of interaction) or $B_{ij}=\infty$ (infinitely strong interaction). This can naturally be realized in reconfigurable atom arrays with mobile ancilla qubits \cite{Bluvstein_2022}, and in static atom arrays using two atomic species \cite{PhysRevA.92.042710,Singh_2023,Bernien,Ireland_2024,saffman-dual} or using local atom-resolved excitation \cite{graham2022multi}. Stabilizers are implemented using four time-optimal CZ gates (up to Hadamards) \cite{Jandura_2022}, with ancilla measurements after each round, and $d$ rounds of syndrome extraction to ensure robustness to measurement errors. Spurious single-qubit $Z$ rotations introduced by the time-optimal CZ gates are compensated at the end of the cycle (see Supp. Material). 

Within this framework, we analyze the dominant error mechanisms occurring before measurements in the limit of rapid QEC cycle execution: The implementation of back-to-back gate operations hinders spontaneous relaxation and leakage events, thereby preserving inter-gate coherences that can translate into non-Markovian correlated errors across sequential gates. We then  investigate fault-tolerant mitigation strategies based on loss biasing, to suppress these correlated errors without sacrificing operational speed (i.e., without  inter-gate latency).

To illustrate loss biasing, we consider an array of trapped neutral atoms, where neighboring atoms can be selectively addressed to implement a symmetric Rydberg blockade gate. Imperfections in the gate trajectory — both coherent (e.g. rotation errors, phase noise, crosstalk) and incoherent (e.g., blackbody transitions, spontaneous decay) — may leave residual population in Rydberg states. These unwanted excitations are removed via rapid ionization and subsequent removal of the remaining electron-ion pair using a small electric field, effectively biasing the gate error toward atom loss. In the case of rare-earth(-like) atoms, this can be done using laser-induced auto-ionization (AI) in which the core electron of the Rydberg atom is resonantly excited to a low-lying $P$-state leading to fast (picosecond-scale) decay to a free-electron-ion pair \cite{cooke1978doubly} [see also level diagram in Fig.~\ref{fig:scalings}(a)]. The speed of Rydberg population removal is therefore determined by the Rabi rate of the coupling laser, which may be made arbitrarily high due to the strong dipole nature of the core electron transition. AI has been demonstrated with near-unity fidelity for reading out Rydberg excitation in Sr on $\mu$s scale \cite{madjarov2020high} (limited by the available laser intensity). In a related application, stroboscopic AI pulses have been used to remove Rydberg contamination and enhance lifetimes in Rydberg-dressed interacting many-body systems \cite{PhysRevLett.134.133201} where — similar to correlated errors in quantum computing — residual Rydberg population can disrupt coherent evolution.

The efficiency of loss biasing can be summarized by a catch-all success probability $p_\text{depl.}\leq1$ of leaving no residual Rydberg population prior to gate execution. While the precise number depends on the detailed mechanics of the gate (e.g. qubit encoding, Rabi rate, excitation trajectory, properties of the Rydberg state, relative strengths of decay channels, etc) our loss biasing analysis is generally relevant for quantum computing architectures employing ground-state, metastable, or optical encoding in alkaline-earth(-like) atoms \cite{PhysRevX.13.041035,zhang2025leveragingerasureerrorslogical,gj3b-5ngl}. 

\begin{figure}[t!]
    \centering
    \includegraphics[width=1.0\linewidth]{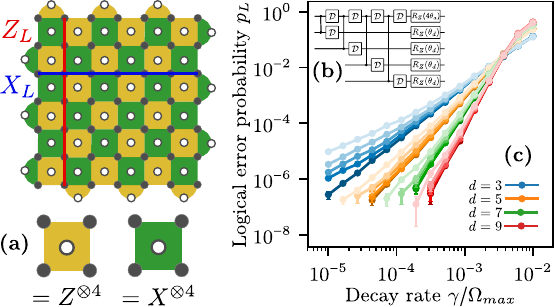}
    \caption{(a) Rotated surface code, stabilizers and logical operators. (b) Stabilizer readout circuit simulated with different success probability of not finding an atom in the Rydberg state before a CZ gate. (c) Surface code logical error probability versus Rydberg decay rate, $\gamma/\Omega_\text{max}$ with time-optimal laser pulses. Data are for $100,90,75,50,0\%$ success probability that an atom is not in the Rydberg state immediately before a the next gate (from dark to light color shading).
    }
    \label{fig:figure1}
\end{figure}

We simulate the open dynamics of the $5$-atom surface code plaquette by generating trajectories from microscopic gate unitaries interleaved with decay events. Decays to $\ket{0}$ or $\ket{1}$ (with equal branching ratio) are modeled by Lindblad operators $L_i^{(q)} = \sqrt{\gamma/2}\ket{q_i}\bra{r_i}$ for $i \in {0,...,4}$ and $q\in {0,1}$, where $\gamma$ is the Rydberg decay rate [Fig.~\ref{fig:scalings}(a)]. Decays to long-lived non-computational states and atom losses imply subsequent CZ gates act trivially, and thus effectively behave as decays to $\ket{0}$ within a single QEC cycle (i.e., until a Hadamard gate is applied). Accordingly, we model Rydberg decay as an amplitude damping channel $\mathcal{D}(\rho) = \Pi \rho \Pi + \bra{r}\rho\ket{r}\Pi/2$, with $\Pi = \ket{0}\bra{0}+\ket{1}\bra{1}$. Inter-gate autoionization is simply modeled as a forced decay to $\ket{0}$ (because qubit replenishment is delayed at the end of the QEC cycle) and systematically vary the probability, $p_\text{depl.}$, of incomplete Rydberg state depletion prior to each gate operation arising from finite auto-ionization efficiency. For instance, a depletion probability of $1$ ($0$) implies that all (none) of the residual Rydberg population is removed before subsequent CZ gates. From these trajectories, we construct an effective Pauli error model via randomized compiling \cite{PhysRevA.88.012314} (see Supp. Material), following \cite{jandura2024surfacecodestabilizermeasurements}. We then simulate the full surface code using the Clifford simulator \texttt{Stim} \cite{Gidney_2021}, measuring $X$ and $Z$ stabilizers separately.

We first focus on characterizing {\it non-Markovian} noise in tightly scheduled QEC cycles: When Rydberg excitations propagate through a CZ gate, excitation hopping mediated by the time‑optimal pulse \cite{jandura2024surfacecodestabilizermeasurements} of the form $\ket{r_1}\bra{1_1}\cdot \text{C}_1\text{Z}_2=\text{C}_1\text{Z}_2\cdot E_1 \cdot\ket{r_2}\bra{1_2}$, with $E_i$ Pauli error on qubit $i$ (see Supp. Material), induces non‑Markovian -- i.e. {\it temporally correlated} --, structures in the effective Pauli channel.  This occurs because consecutive gates preserve residual coherences between operations, precluding the relaxation of excitations and thereby inducing non-Markovian memory between gate errors. Upon Pauli twirling, this non-Markovianity is revealed as {\it spatially correlated} (hook-like) multi-qubit error strings produced by single decay events that violate fault‑tolerance, as shown in detail below.
We note that these non-Markovian correlations in the effective Pauli model originate from the continuous evolution spanning subsequent gates, even though the underlying microscopic dynamics 
remains Markovian.

\begin{figure}
    \centering
    \includegraphics[width=1.0\linewidth]{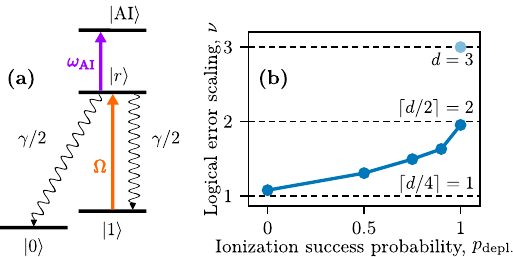}
    \caption{(a) Schematic level structure. (b) Logical error scaling $\nu$ such that $p_L\propto \gamma^\nu$ versus success probability, $p_\text{depl.}$, of inter-gate ionization for distance-$3$ surface codes. Back-to-back gate schedules produce a non-fault-tolerant scaling $\lceil d/4\rceil$, while perfect inter-gate ionization restores the fault-tolerant Pauli scaling of $\lceil d/2\rceil$ for Pauli errors. If losses are properly handled in software with a loss-aware decoder, the optimal erasure-like scaling of $d$ can be achieved (light blue). Error bars fall within the marker size.}
    \label{fig:scalings}
\end{figure}

To probe the link between non-Markovian noise and the breakdown of fault tolerance, we analyze surface code stabilizer circuits with autoionization applied after each CZ gate on both data and ancilla atoms, as shown in Fig.~\ref{fig:figure1}(b). After every CZ, the participating atoms undergo an autoionization pulse modeled as an amplitude damping channel $\mathcal{D}(\rho)$ with decay probability $\gamma\propto p_\text{depl.}$. We consider percentage success probabilities of $100\%$, $90\%$, $75\%$, $50\%$, and $0\%$ (no autoionization).

Monte Carlo results for code distances $d=3$–$9$ are shown in Fig.~\ref{fig:figure1}(c), with color shading from dark ($100\%$) to light ($0\%$) describing the different success probabilities. Syndrome data are decoded using \texttt{pymatching} \cite{Higgott2025sparseblossom}. In this plot, the logical error probability exhibits different asymptotic scalings depending on the ionization probability; as an example, we summarize the asymptotic fits for $d=3$ surface codes in Fig.~\ref{fig:scalings}(b). 
These results show two interesting effects: first, near-perfect autoionization ($100\%$) gives $\nu\approx\lceil d/2\rceil$, consistent with fault-tolerant scaling, whereas the absence of ionization ($0\%$) yields $\nu\approx\lceil d/4\rceil$, indicating non-fault-tolerant behavior due to the buildup of correlated errors. These different asymptotic scalings result in a gain in QEC performance exceeding one order of magnitude already for $\gamma\sim10^{-4}\Omega_\text{max}$ [see Fig.~\ref{fig:scalings}(c)].
These results are thus interesting as they challenge the common assumption that \emph{faster is always better} for minimizing error rates, which in current experiments are dominated by spontaneous emission and Rydberg leakage errors. In fact, rapid implementations can still produce a small number of strongly correlated errors that degrade fault tolerance. Second, we note that in the range $\gamma\sim 10^{-3}\Omega_\text{max}$ 
experimentally accessible today, simulations in Fig.~\ref{fig:figure1}(c) show minimal difference in logical performance for ionization success probabilities $100,90,75\%$. 
These results suggest that, in the near term, partial inter-gate Rydberg ionization — sufficient to convert most, though not all, of the Rydberg population into loss — can achieve near-optimal logical error probabilities for Pauli errors, despite the absence of asymptotic fault tolerance, that is, despite an asymptotic $\nu<\lceil d/2\rceil$.

In the Supp. Material, we further present surface code simulations with \emph{perfect} inter-gate ionization applied selectively within the stabilizer readout circuit, rather than after every gate. We consider ionization applied: (i) only on ancilla qubits after each gate, (ii) only on data qubits, (iii) once midway through the circuit, and (iv) twice, after the first and third CZ gates. This allows us to directly test whether the stringent requirement of ionizing all atoms after every gate can be relaxed.
While this requirement can be readily met in alkaline-earth(-like) platforms, it poses a significant challenge for alkali atoms, where autoionization is not available and alternatives such as pulsed field ionization \cite{PhysRevLett.70.1236} are substantially more demanding. Remarkably, we find that ionizing only the ancilla after each gate — assuming data–ancilla blockade — is already sufficient to preserve fault tolerance, whereas all other schedules fail to do so.
This result identifies a minimal and experimentally viable ionization strategy,  directly applicable to static dual-species architectures \cite{PhysRevA.92.042710,Singh_2023,Bernien,Ireland_2024,saffman-dual}, where species-selective global control enables independent addressing of ancilla qubits. A detailed discussion of implications for single- and dual-species alkali platforms is provided in the Supp. Material.

In order to assess a quantitative relation between noise non-Markovianity and lack of fault tolerance, we now consider the error distribution of a single surface code plaquette. The effective twirled five-qubit Pauli channel extracted from the microscopic model is of the form $\mathcal{E}(\rho)=\sum_Q\lambda_QQ\rho Q$, where $Q\in\{I,X,Z,Y\}^{\otimes5}$ is any five-qubit Pauli string and $\lambda_Q\equiv\lambda_Q(\gamma)$ its probability to occur, which depends on $\gamma$. We fit the $\lambda_Q(\gamma)$ with the empirical functional dependence $A\gamma^n+B\gamma^{(n+1)}$ with $A,B,n$ constants. We count the number of error keys with $n=1$ containing pairs of $\{Z,Y\}^{\otimes2}$ ($\{X,Y\}^{\otimes2}$) Pauli errors on data qubits that align with the surface logical operator $Z_L$ ($X_L$), for $X$- and $Z$-stabilizers. These correspond to correlated hook-like errors produced by a single fault (decay event) which can trigger a logical error and hence are symptoms of decay-induced non-Markovianity. 
We plot in Fig.~\ref{fig:figure2}(a)--(b) the occurrences of these hook error strings using ionization on all atoms with different success probabilities and only at selected circuit location with perfect success probability. We find that imposing perfect ionization after each gate -- either on all qubits or only on the ancilla qubit -- results in no hook error string, consistently with the observed fault-tolerant scaling of the logical error probability. Instead, all other protocols suffer from these errors and are not fault-tolerant. Along with their occurrences, in Fig.~\ref{fig:figure2}(c) we plot the amplitudes of these errors for all different gate schedules that we study. Interestingly, we find that permitting $75\%$ or $50\%$ ionization success probabilities after each gate yields logical error probabilities comparable to those obtained by enforcing perfect ionization only at mid‑cycle or twice within the stabilizer measurement round (see Supp. Material). Nevertheless, the former schedules suffer from a substantially larger number of hook error strings than the latter. 

The analysis above points to two distinct physical mechanisms that enhance QEC performance. Inter-gate imperfect ionization diminishes the noise amplitude, reducing the probability of each potential error string. Instead, perfect ionization at selected circuit locations suppresses the effective noise entropy by disrupting temporal correlations, so that a single decay event can generate only a limited set of hook error strings with large amplitude, as visible in Fig.~\ref{fig:figure2}(c) (see Supp. Material for a more detailed discussion). This analysis shows that the QEC performance of a surface code depends both on the number of such errors and on their amplitude, and that in the near term logical error probabilities can be decreased by acting to reduce either of these quantities separately.

All these results point to loss-biasing as a viable strategy to improve QEC performance. 
To prevent propagation across QEC cycles, atom losses can be addressed using leakage detection units and replenishment of the qubit array. Theoretical studies demonstrate that heralded \cite{Perrin_2025,ycwc-3myc} or highly correlated \cite{perrin2026correlatedatomlossresource} losses can be mapped to delayed erasures, yielding a logical error scaling that approaches the optimal $\sim d$ scaling characteristic of erasure channels. We note that, by focusing on the intra-cycle correction of Rydberg hopping errors to restore the fault-tolerant $\lceil d/2\rceil$ scaling for Pauli errors, our protocol ensures that this optimal $\sim d$ scaling is successfully recovered when losses are mitigated in software -- a non-fault tolerant syndrome extraction with scaling of $\lceil d/4\rceil$ is instead expected to yield $\lceil d/2\rceil$ for delayed erasures. Furthermore, our loss-biasing scheme easily accommodates correlated losses, which experimental evidence \cite{senoo2025highfidelityentanglementcoherentmultiqubit} shows are significantly more probable than independent two-atom losses. Specifically, if one atom decays during a gate, its interacting partner acquires a residual Rydberg population, increasing its probability of subsequent decay or of remaining in a Rydberg state susceptible to ionization. Finally, the loss-biasing protocol is intrinsically compatible with loss-heralding techniques based on ion detection. For example, one can image the ions mid-circuit using fluorescence imaging in a tens of $\mu$s timescale \cite{Wu_2022}. Alternatively, ion detection can be achieved with spatial resolution using a microchannel plate, or via a channeltron; the latter, while lacking spatial resolution, provides exceptional timing precision  ($\sim1$ ns) and high detection efficiency ($\sim0.9$) \cite{Hofmann,veit21}, useful for example, when syndrome extraction gates are shined sequentially across stabilizers. 
Even more recent theoretical advances \cite{liu2026achievingoptimaldistanceatomlosscorrection} show that atom loss errors can reach the optimal $\sim d$ scaling even when neither heralded nor correlated, provided they are managed and replenished using a decoder that exploits their intrinsic non-Markovianity to assign uniquely identifiable syndromes. Unlike Pauli errors, which propagate linearly, losses behave non-linearly by preventing subsequent gates. These results indicate that atom loss can match the effective distance of erasure errors, at the only cost of increased classical decoding complexity, further supporting loss biasing as a pathway to fast, sub-millisecond fault-tolerant quantum error correction. 
\begin{figure}
    \centering
    \includegraphics[width=1.0\linewidth]{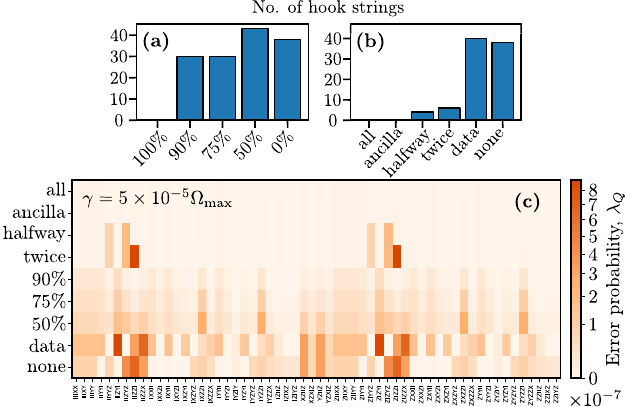}
    \caption{Occurrences of hook error strings caused by single decay events that degrade fault tolerance. Number hook errors for (a) ionization enforced after each gate with success probabilities ranging from $0\%$ to $100\%$ and (b) perfect ionization only enforced at selected circuit locations (see main text). (c) Probability of each hook error string at fixed decay rate $\gamma=5\times10^{-5}\Omega_\text{max}$ (far below the surface code threshold). }
    \label{fig:figure2}
\end{figure}

In conclusion, in this work, we have analyzed temporal correlations by Rydberg excitation hopping affecting  surface code stabilizer measurement cycles. Our analysis indicates that ionization promotes relaxation from the Rydberg manifold to the ground state and suppresses inter‑gate coherences, thereby rendering the resulting Pauli noise effectively Markovian. By contrast, fully and partially back‑to‑back gates retain residual coherences that project onto temporally correlated, non‑Markovian Pauli errors with spatial signatures. Crucially, only gate schedules enforcing perfect ionization after each gate, either on all atoms or only on the ancilla (at least in the assumption of data-ancilla blockade), result in purely Markovian Pauli noise and succeed to maintain fault tolerance. We have also shown that relaxation is compatible with qubit loss and suggested fast inter-gate loss conversion of spurious Rydberg excitations opposed to latency to preserve fault tolerance and improve QEC performance without sacrificing speed. This systematic inter-gate loss-conversion practically biases the noise towards atom loss errors, which can be handled in software to greatly improve both threshold and logical error probabilities, achieving the optimal scaling of $\sim d$. 
All these findings highlight the focus on extracting hardware-specific error information, while concurrently reducing QEC cycle times in neutral atom platforms.

Although our study focuses on surface codes, the conclusions extend to other syndrome extraction schemes (notably those employing multiqubit gates \cite{three-qubit,3qubit-old}) and QEC codes, including quantum Low-Density Parity-Check (LDPC) codes \cite{Bravyi_2024,xu2023constantoverheadfaulttolerantquantumcomputation,Pecorari_2025,poole2025,mgkt-ctv8,pecorari2025addressablegatebasedlogicalcomputation}. The long-range connectivity intrinsic to these codes typically requires longer operational times (e.g., due to qubit shuttling or static long-range connections) which can favor relaxation to the ground state of idling qubits. The mechanisms and mitigation principles identified here are likewise relevant beyond QEC, e.g., for hybrid analog–digital or trotterized quantum simulations, and are expected to generalize to any quantum computing platform whose gates involve or are vulnerable to leakage into non‑computational states.

\emph{Acknowledgments -- We thank Asier
Piñeiro Orioli, Hugo Perrin, and Shannon Whitlock for useful discussions. We acknowledge funding from the European Union’s Horizon 2020 research and innovation programme under the Horizon Europe programme HORIZON-CL4-2021-DIGITAL-EMERGING-01-30 via the project 101070144 (EuRyQa) and from the French National Research Agency under the Investments of the Future Program projects ANR-21-ESRE-0032 (aQCess), ANR-22-CE47-0013-02 (CLIMAQS), ANR-17-EURE-0024 (QMat), and ANR-22-CMAS-0001 France 2030 (QuanTEdu-France). }

\bibliography{references}

\onecolumngrid
\clearpage
\begin{center}
\textbf{\large Supplemental Material for ``Loss-biased fault-tolerant quantum error correction"}

\vspace{0.5cm}

Laura Pecorari$^1$, Gavin K. Brennen$^{2,3}$, Stanimir Kondov$^1$, and Guido Pupillo$^1$

\emph{$^1$University of Strasbourg and CNRS, CESQ and ISIS (UMR 7006), aQCess, 67000 Strasbourg, France}

\emph{$^2$Center for Engineered Quantum Systems, School of Mathematical and Physical Sciences, Macquarie University, 2109 NSW, Australia}

\emph{$^3$BTQ Technologies, 16-104 555 Burrard Street, Vancouver, British Columbia, Canada V7X 1M8}

{\small (Dated: \today)}
\end{center}
\maketitle
\setcounter{equation}{0}
\setcounter{figure}{0}
\setcounter{table}{0}
\makeatletter
\renewcommand{\theequation}{S\arabic{equation}}
\renewcommand{\thefigure}{S\arabic{figure}}
\twocolumngrid

\section{S1. Time-optimal pulse and Rydberg excitation hopping}
We now recall the finding of Ref.~\cite{jandura2024surfacecodestabilizermeasurements} on why the time-optimal gate pulse mediates Rydberg excitation hopping. The time-optimal pulse is symmetric, i.e. both atoms involved in the CZ gates are acted on by the same pulse with complex Rabi frequency $\Omega(t)$. We restrict to the $\text{span}\{\ket{01},\ket{0r},\ket{11},\ket{W}\}$ space with $\ket{W_\pm} = (\ket{1r}\pm\ket{r1})/\sqrt{2}$. This is possible because $\ket{00}$ evolves trivially, $\ket{rr}$ is never populated in the infinite blockade limit, and $\ket{10}$, $\ket{r0}$ evolve symmetrically to $\ket{01}$, $\ket{0r}$. Thus, the two-qubit gate Hamiltonian reads $H_{2q}=H_1+H_2$ with 
\begin{align*}
\begin{split}
    H_1&=\frac{\Omega(t)}{2}\ket{0r}\bra{01}+\mathrm{h.c.} \\
H_2&=\frac{\sqrt{2}\Omega(t)}{2}\ket{W_+}\bra{11}+\mathrm{h.c.}\,.    
\end{split}
\end{align*}
The optimal pulse is obtained by integrating the Schr\"odinger equation for $H_1$ and $H_2$, and it implements a unitary $V_\text{CZ}$ equivalent to a CZ gate followed by a single-qubit rotation $R_Z(\theta)$ by some angle $\theta$. Thus, in the two-atom basis, $V_\text{CZ}\ket{01}=e^{i\theta}\ket{01}$ and $V_\text{CZ}\ket{11}=e^{2i\theta}\ket{11}$. Since $\text{tr}(H_1)=\text{tr}(H_2)=0$, it follows that $V_\text{CZ}$ must have unitary determinant restricted to the two corresponding subspaces, i.e. $\{\ket{01},\ket{0r}\}$ and $\{\ket{11},\ket{W_+}\}$. Hence, it must be $V_\text{CZ}\ket{0r}=-e^{-i\theta}\ket{0r}$, $V_\text{CZ}\ket{W_+}=-e^{-2i\theta}\ket{W_+}$, and $V_\text{CZ}\ket{W_-}=\ket{W_-}$ due to the pulse symmetry. Thus, it follows that $V_\text{CZ}\ket{1r}=e^{-i\theta}\left(i\text{sin}(\theta)\ket{1r}-\text{cos}(\theta)\ket{r1}\right)$ and $V_\text{CZ}\ket{r1}=e^{-i\theta}\left(i\text{sin}(\theta)\ket{r1}-\text{cos}(\theta)\ket{1r}\right)$, revealing that the time-optimal pulse mediates Rydberg excitation hopping between the two atoms involved in the CZ gate.

\section{S2. Details on Rydberg decay simulations}
We model each atom as a three level system, with computational basis states $\ket{0}$, $\ket{1}$ and auxiliary Rydberg state $\ket{r}$. The dynamics is governed by a Lindblad master equation  $\dot{\rho} = -i[H,\rho] + \sum_{i,q} L_i^{(q)}\rho {L_i^{(q)}}^\dag -\{{L_i^{(q)}}^\dag L_i^{(q)}, \rho\}/2$ with Hamiltonian ($\hbar=1$) $H =\sum_{i=0}^4\Delta E_i\ket{1_i}\bra{1_i}+\sum_{ij=0}^{4} B_{ij}\ket{r_ir_j}\bra{r_ir_j} + \sum_{i=0}^{4} \frac{\Omega_i(t)}{2}\ket{r_i}\bra{1_i} + \mathrm{h.c.}$ and collapse operators $L_i^{(q)} = \sqrt{\gamma/2}\ket{q_i}\bra{r_i}$ for $i \in \{0,...,4\}$ and $q\in \{0,1\}$ (see main text for notation). We first initialize the ancilla atom at the center of the plaquette to state $\ket{+}=\left(\ket{0}+\ket{1}\right)/\sqrt{2}$ and apply time-optimal symmetric laser pulses with complex Rabi frequency $\Omega(t)$ on the ancilla and the four data atoms, respectively. The pulses are chosen such that, without decay ($\gamma=0$), we implement a unitary $U$ corresponding to four controlled-Z (CZ) gates between the ancilla and each data atom followed by single-qubit rotations $R_Z(\theta)=\exp(i\theta \ket{1}\bra{1})$ and $R_Z(4\theta)$ on data and ancilla atoms, respectively. Single-qubit gates are then applied to each atom to compensate for the single-qubit rotation induced by $U$, namely we apply $R_Z(-4\theta)$ and $R_Z(-\theta)$ on ancilla and each data atom, respectively. We simulate the single-plaquette noisy stabilizer measurement as four ideal CZ gates between the ancilla and the data qubits followed
by a 5-qubit noise channel $\mathcal{E}_\text{err}(\rho)$ and the measurement of the ancilla qubit. We use randomized compiling and insert random Pauli
gates $P$ and their Clifford conjugate gates $P' = U P U^\dagger$ around $U$ to re-write $\mathcal{E}_\text{err}(\rho)$ as a diagonal Pauli channel of the form $\mathcal{E}_\text{err}(\rho)=\sum_Q\lambda_QQ\rho Q$, with $Q$ any $5$-qubit Pauli string and $\lambda_Q$ its associated probability. 

We compute the Pauli‑twirled error probabilities $\lambda_Q$ by simulating an operator‑sum, Trotterized approximation to the Lindblad dynamics on the qutrit space $\{\ket{0},\ket{1},\ket{r}\}$. The protocol is decomposed into short time slices; on each slice the coherent part is applied exactly by exponentiating the two-qubit gate Hamiltonian $H_{2q}=\frac{\Omega(t)}{2}\ket{0r}\bra{01}+\frac{\sqrt{2}\Omega(t)}{2}\ket{W_+}\bra{11}+\mathrm{h.c.}$ with $\ket{W_+} = (\ket{1r}+\ket{r1})/\sqrt{2}$, and dissipation is applied as instantaneous amplitude damping maps that transfer population out of $\ket{r}$ back to the ground state with instantaneous probability $\gamma dt$. Inter-gate perfect ionization is simulated by forcing all the population to decay to the $\ket{0}$ state with unitary probability. We note that this procedure converges to the continuous Lindblad solution in the limit $dt\rightarrow0$. After propagating any $5$-qubit Pauli string through the full sequence, we construct the Pauli transfer matrix and extract the diagonal Pauli‑twirled probabilities $\lambda_Q$. 

Finally, we note that we focus on error correlations from leakage \emph{within} individual QEC cycles and therefore our simulations do not include leakage reduction units between cycles (meaning that qubits in $\ket{0}$ are mapped to $\ket{+}$ upon encountering a Hadamard gate).

\section{S3. Error probabilities, decay events and fault tolerance}
After having calculated the twirled error probabilities $\lambda_Q$ for any $5$-qubit Pauli string $Q$, we extract their scaling as a function of the decay rate $\gamma$. We fit the $\lambda_Q$ with the functional dependence $A\gamma^n +B\gamma^{(n+1)}$ with $A$, $B$ constants, and $n$ counting the number of decay events producing a given error string. This allows us to filter out all the $5$-qubit Pauli strings $Q$ that can cause a logical error while being produced by a single decay event ($n=1$) and are therefore responsible for degrading fault tolerance. We recall that fault tolerance requires that a single fault on a data qubit does not propagate to more than one data qubit or, if it propagates to multiple data qubits, it does not align with a logical operator in the same basis. For example, if a single-qubit fault $E$ propagates as $E\rightarrow ZZ$, the pair of correlated $ZZ$ errors must not align with the $Z_L$ logical operator.

\section{S4. Error distribution, noise amplitude and noise entropy}
To probe fault‑tolerance and characterize non-Markovian correlated errors, we fit the twirled Pauli probabilities $\lambda_Q$ as described in the previous section and restrict the analysis to the Pauli strings with power‑law exponent $n = 1$ (i.e., errors that scale linearly with $\gamma$). In particular, we focus on error strings that trigger a logical error, i.e. that contain pairs of $\{Z,Y\}^{\otimes2}$ ($\{X,Y\}^{\otimes2}$) errors on data qubits which align with the surface logical operator $Z_L$ ($X_L$), for both $X$- and $Z$-stabilizers. In the simulated surface codes, $Z_L$ logical operators are vertical, while $X_L$ logical operators are horizontal [Fig.~1(a) main text]. $5$-qubit Pauli error strings are ordered with the ancilla qubit in the leading position, and the remaining data qubits follow the rotated surface‑code fault‑tolerant ordering to ensure robustness against hook errors \cite{tomita}. 

In the main text and in main Fig.~3(a) and (b), we have counted the occurrences of such errors for different protocols, showing that protocols that offer fault-tolerant logical error scaling consistently do not suffer from any of these errors. Notably, these were protocols allowing full relaxation after each gate, either on both qubits involved in the CZ gate or only on the ancilla qubit. We have also commented that, by comparing partial and selective relaxation, gate schedules with comparable logical error probabilities can suffer from both large and small numbers of these hook error strings. This observation pointed to two distinct physical mechanisms that enhance QEC
performance: reduction of noise amplitude by partial relaxation and reduction of noise entropy by selective relaxation. To support this statement, together with the occurrences of these errors, here we study their probability to occur at fixed far-below threshold decay rate, say $\gamma=5\times10^{-5}\Omega_\text{max}$. Results are shown in Fig.~3(c) (main text). Schedulings enforcing partial $50-90\%$ inter-gate relaxation allow for several hook error strings (large number of error configurations, large entropy), but the amplitude of these strings gets smaller as the inter-gate relaxation time increases from $50\%$ to $90\%$ (reduction of noise amplitude). For full inter-gate relaxation, these amplitudes vanish, no hook error string occurs and fault tolerance is preserved. Protocols enforcing full relaxation only at half cycle or after the first and third CZ gates allow for fewer hook error strings (small number of configurations, small entropy) but their probability to occur is large (large noise amplitude). These observations show that the logical error probability is governed by the combined effect of multiple factors, such as the number of distinct Pauli errors that can occur and their individual probabilities. As a result, contributions to logical fidelity are highly nonuniform and a relatively small subset of error configurations typically dominates the logical‑error budget in the deep sub-threshold regime.

\section{S5. Relaxation at selected qubit location}

We now want to investigate whether it is possible to relax the hardware constraint of ionizing all atoms after each gate in the stabilizer measurement circuit. This is relevant for alkali neutral atom platforms, where autoinization is not viable and other more technically challenging techniques need to be adopted to rapidly expel Rydberg excitations from the traps without relying on latency (see next supp. section). 

\begin{figure}
    \centering
    \includegraphics[width=0.99\linewidth]{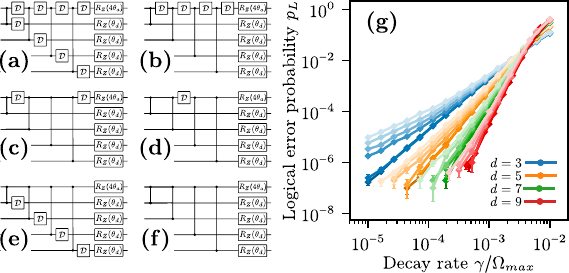}
    \caption{Surface code stabilizer readout circuits with Rydberg ionization (relaxation) inserted at selected locations: (a) on data and ancilla atoms after each gate; (b) only on ancilla; (c) only on ancilla after gates $1$ and $3$ (twice); (d) only on ancilla halfway through the circuit; (e) only on data; (f) none (purely sequential scheme). (g) Surface code logical error probability versus Rydberg decay rate, $\gamma/\Omega_\text{max}$ with time-optimal laser pulses and ionization at selected circuit locations. 
    }
    \label{fig:figure3}
\end{figure}

We consider the case of perfect ionization (relaxation) applied only at specific locations within the surface code stabilizer measurement circuit. In particular, we focus on the following cases, with ionization applied: after each gate on both atoms involved in the gate, only on the ancilla after each gate, only on the ancilla after the first and the third CZ gate (twice), only on the ancilla halfway through the circuit, only on the data atoms after each gate, and on no atom (none), corresponding to the circuits depicted in Fig.~\ref{fig:figure3}(a)--(f). 

The results of Monte Carlo simulations are shown in Fig.~\ref{fig:figure3}(g) for surface code distances from $d = 3$ to $d = 9$; different schedulings are represented with different color shadings, from dark to light, in the same order as they are discussed above. Syndrome information is decoded with \texttt{pymatching}. Plots show different asymptotic scalings of the logical
error rate. Fits of the logical error probabilities of $d=3$ surface codes reveal: $\nu_\text{all}=1.96$, $\nu_\text{ancilla}=1.98$, $\nu_\text{twice}=1.31$, $\nu_\text{halfway}=1.23$, $\nu_\text{data}=1.14$, and $\nu_\text{none}=1.08$. All schedules improve logical error suppression compared to a purely sequential readout without ionization. At the same time, they all fail to maintain fault tolerance except for those imposing ionization-mediate relaxation of all qubits or only of the ancilla qubits after each gate, which show optimal scaling $\nu\approx\lceil d/2\rceil$. This suggests that imposing relaxation only on the ancilla qubits—using independent controls from the data qubits as in dual-species atom arrays—may offer a promising hardware-efficient pathway towards fault tolerance without the need of ionizing all atoms after each gate.

\section{S6. Alkali qubit platforms}
In this section, we show how our findings extend to alkali atom platforms. We discuss alkali-specific mid-circuit ionization strategies for alkali-based erasure conversion and comment separately on single-species and dual-species platforms. For the latter, as found in the above supplemental section, static designs with species-selective ionization represent a practical pathway to realize sub-ms QEC cycles with our loss-biasing protocol.

We start by focusing on single-species alkali atom platforms and consider an array of trapped $^{87}$Rb atoms with qubits encoded in two hyperfine clock states. For stabilizer readout, state-of-the-art experiments utilize a movable ancilla shuttled to each neighboring data atom in $200-400\mu$s per transfer. Optical traps are switched off for $500$ns during each gate to avoid anti‑trapping of the Rydberg state \cite{Bluvstein_2022}. Recent experiments \cite{Bluvstein_2025} identify black body–induced decay to nearby Rydberg $P$ states ($\sim100\mu$s lifetimes) as dominant error source. Gate sequences with $4\mu$s spacing are therefore prone to these errors, whereas circuits implemented via atom rearrangement are not, as they allow time for decay or for ejection when traps are pulsed on \cite{Bluvstein_2025}. Our findings on latency are consistent with these results and further show that reduction in gate fidelity translates into degradation of logical fidelity and fault tolerance.

In contrast, dual-species alkali architectures utilize different atomic species for data and ancilla qubits, permitting in-place syndrome extraction free of measurement crosstalk. While this makes them ideal candidates for rapid QEC cycles, it also makes them vulnerable to the Rydberg hopping errors discussed in this work. As shown in the previous supplementary section, fault tolerance can be preserved in these systems by ionizing exclusively the ancilla species after each gate, a procedure easily implemented using species-selective global control. Because autoionization is fundamentally unsuited for alkali atoms, we identify pulsed field ionization \cite{PhysRevLett.70.1236} as a highly efficient alternative loss-biasing strategy. Nevertheless, despite its generality, this approach remains technically challenging due to the necessity of rapidly switching large electric fields ($\sim1$ kV/cm).

\end{document}